\documentclass[aps,reprint]{revtex4-1}%
\usepackage{amssymb}
\usepackage{amsfonts}
\usepackage{amsmath}
\usepackage{graphicx}%
\begin{document}
\title{The dynamic three-dimensional Anderson localization of optical fields in
active percolating systems}
\author{Gennadiy Burlak}
\affiliation{Centro de Investigaci\'{o}n en Ingenier\'{\i}a y Ciencias Aplicadas,
Universidad Aut\'{o}noma del Estado de Morelos, Cuernavaca, Mor., M\'{e}xico}

\begin{abstract}
We study three-dimensional optical Anderson localization in medium with a
percolating disorder, where the percolating clusters are filled by the light
nanoemitters in the excited state. The peculiarity of situation is that in
such materials the field clusters join a fractal radiating system where the
light is both emitted and scattered by the inhomogeneity of clusters. Our
numerical FDTD simulations found that in this nonlinear compound the number of
localized field bunches drastically increases after the lasing starts. At
longer times the bunches leave nonlinear percolating structure and are
radiated from the medium. Monitoring the output of system allows to observe
experimentally such dynamic localized bunches in three-dimensional setup.

\end{abstract}
\maketitle

\textit{Introduction}. Disordered photonic materials can diffuse and localize
light through random multiple scattering that leads to formation of the
electromagnetic modes depending on the structural correlations, scattering
strength, and dimensionality of the system \cite{Riboli:2014},
\cite{Vinck:2011}, \cite{Sheng:2010}, \cite{Wang:2011},
\cite{jendrzejewski:2012}, \cite{Segev:2013a}, \cite{Wiersma:2013a}. The
Anderson localization was predicted as a non-interacting linear interference
effect \cite{anderson:1958}. However, in real systems the non-negligible
interactions between light and medium can take place. Therefore important
aspect of the optical localization is the interplay between nonlinear
interactions and linear Anderson effect \cite{Segev:2013a}. Nonlinear
interactions appear in optics, due to nonlinear responses of a disordered
medium that normally gives rise to indirect interactions between the photons
through various mechanisms. In case of classical waves the localization can be
interpreted as interference between the various amplitudes associated with the
scattering paths of optical waves propagating among the diffusers. The study
for Anderson transition in 3-D optical systems still has not been conclusive
despite considerable efforts. The localization transition may be difficult to
reach for the light waves due to various effects in dense disordered media
required to achieve strong scattering [see \cite{Skipetrov:2016a} and
references therein]. The experimental observation of optical Anderson
localization \cite{Sanli:2009a} just below the Anderson transition in 3D
disordered medium shows strong fluctuations of the wave function that leads to
nontrivial length-scale dependence of the intensity distribution (multifractality).

Here we study the optical Anderson localization in 3D percolating disorder,
where the percolating clusters are filled by the light nanoemitters in the
excited state. The peculiarity of situation is that in such materials the
field clusters join a fractal nonlinear radiating system where the light is
both emitted and scattered by the inhomogeneity of clusters. Such system may
be suggested as an extension to 3D optical case the localization released in
one-dimensional waveguide in the presence of a controlled disorder
\cite{Billy:2008a}. In 3D disordered percolating systems the optical transport
was observed \cite{Burlak:2009a} and also it is found that the random lasing
assisted by nanoemitters incorporated into such a disordered structure can
occur \cite{burlak:2015}. The system with spanning clusters produces a global
percolation that results in qualitatively modification of its spatial
properties. One can argues that already in a vicinity of the percolating phase
transition the fractal dimension of such system $D_{H}\simeq2.54$ considerably
differs from the dimension of the embedded space $D=3$ (multifractality). The
crucial question is whether the optical Anderson localization can be achieved
for a non-integer dimension case with a fractal (Hausdorff) dimension of
$D_{H}<3$, where the strong randomness for properties of the system is
expected. In this paper we show that the dynamic field localization can arise
in such 3D active fractal percolation system. To the best of our knowledge,
the dynamic 3D Anderson localization assisted by fractal disordered
inhomogeneity is discussed here for the first time.

\begin{figure}[ptb]%
\includegraphics[width=9cm]%
{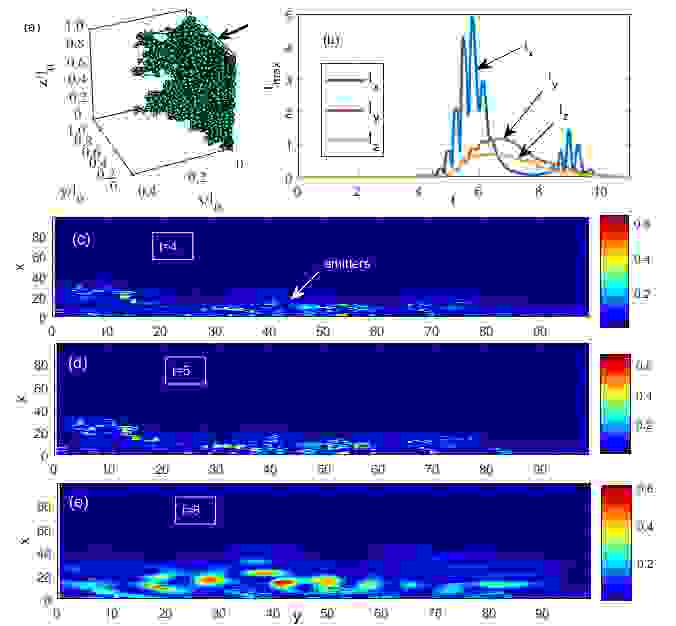}%
\caption{(Color online.) (a) Typical spatial structure of the incipient
percolating cluster near the percolation threshold in the cube $l_{0}\times
l_{0}\times l_{0}$. (b) The dynamics of lasing (the critical lasing time is
$t_{c}=5$), where the field fluxes $I_{xyz}$ in the $xyz$ -directions of the
system are displayed; (c) The normalized fields in the central intersection
$xy$ of 3D system for various times closely to the critical time at: (c)
$t=4<t_{c}$, and (d) $t=5=t_{c}$. In such situation only narrow point-like
fields generated by random emitters are observed. For longer time at
$t=6>t_{c}$ (e) the field bunches having smooth localized shape and large
amplitudes arise already beyond the emitter peaks. Hence the lasing in
percolating disordered system will stimulate the localization of emitted
field.}%
\label{Pic_Fig1}%
\end{figure}

\textit{Basic equations.} The percolation normally refers to the leakage of
fluid or gas through the porous materials. In this paper we consider that the
percolating clusters are filled by gaseous light nanoemitters in the excited
state. A typical incipient percolation cluster has a dendrite shape. From Fig.
\ref{Pic_Fig1} (a) we observe that such a system consists of
two essentially different areas. In the first region (in the vicinity of the
entrance shown by arrow) there is a strong concentration of percolation
clusters, which may lead to accumulation of the nonlinear field effects and
the lasing. In the other part, the concentration of percolation clusters is
small, so the disorder of medium is lower. We study the emission of
electromagnetic energy from a cubical sample $(x,y,z)\in\lbrack0,l_{0}]$ in 3D
medium with a percolating disorder, where the percolating clusters are filled
by the light nanoemitters in the excited state. The output flux of energy can
be written as $I=\oint_{S}(\mathbf{K}\cdot\mathbf{n})dS=I_{x}+I_{y}+I_{z}$,
where $\mathbf{K}$ is the Pointing vector, $\mathbf{n}$ is the normal unit
vector to the surface $S$ of cube, and $I_{x,y,z}$ indicate the fluxes from
two faces of the cube perpendicular to a particular direction. To find the
emission from the system we solve numerically the following equations that
couple the polarization density $\mathbf{P}$, the electric field $\mathbf{E}$,
and occupations of the levels of emitters $N_{i}$ \cite{Siegman:1986} $\ $%

\begin{equation}%
\begin{tabular}
[c]{l}%
$\frac{\partial^{2}\mathbf{P}}{\partial t^{2}}+\Delta\omega_{a}\frac
{\partial\mathbf{P}}{\partial t}+{\omega_{a}^{2}}\mathbf{P}=\frac
{6\pi\varepsilon_{0}c^{3}}{\tau_{21}\omega_{a}^{2}}(N_{1}-N_{2})\mathbf{E,}$\\
$\frac{\partial N_{0,3}}{\partial t}=\mp A_{r}N_{0}\pm\frac{N_{1,3}}%
{\tau_{(10),(32)}},$\\
$\frac{\partial N_{1,2}}{\partial t}=\frac{N_{2,1}(t)}{\tau_{(21),(32)}}%
\mp\frac{(\mathbf{j}\cdot\mathbf{E})}{\hbar\omega_{a}}-\frac{N_{1,2}}%
{\tau_{(10),(3.2)}},$%
\end{tabular}
\ \ \ \label{Eqs}%
\end{equation}
here $\Delta\omega_{a}=\tau_{21}^{-1}+2T_{2}^{-1}$, where $T_{2}$ is the mean
time between dephasing events, $\tau_{21}$ is the decay time from the second
atomic level to the first one, and $\omega_{a}$ is the frequency of radiation.
The electric and magnetic fields, $\mathbf{E}$ and $\mathbf{H}$, and the
current $\mathbf{j}=\partial\mathbf{P}/\partial t$ are found from the Maxwell
equations, together with the equations for the densities $N_{i}(\mathbf{r},t)$
of atoms residing in $i-$th level, see \cite{Soukoulis:2000} and references
therein. An external source excites emitters from the ground level ($i=0$) to
third level at a certain rate $A_{r}$ , which is proportional to the pumping
intensity in experiments. After a short lifetime $\tau_{32}$, the emitters
transfer nonradiatively to the second level. The second level and the first
level are the upper and the lower lasing levels, respectively. Emitters can
decay from the upper to the lower level by both spontaneous and stimulated
emission, and $(\mathbf{j}\cdot\mathbf{E})/\hbar\omega_{a}$ is the stimulated
radiation rate. Finally, emitters can decay nonradiatively from the first
level back to the ground level. The lifetimes and energies of upper and lower
lasing levels are $\tau_{21}$, $E_{2}$ and $\tau_{10}$, $E_{1}$, respectively.
The individual frequency of radiation of each emitter is then $\omega
_{a}=\left(  E_{2}-E_{1}\right)  /\hbar$, and $\hbar$ is Planck's constant.

\textit{Numerics. }In our calculations we considered the gain medium with
parameters close $GaN$ powder, similar to Ref. \cite{expt}. The lasing
frequency $\omega_{a}$ is $2\pi\times3\times10^{13}\,Hz$, the lifetimes are
$\tau_{32}=0.3\,ps$, $\tau_{10}=1.6\,ps$, $\tau_{21}=16.6\,ps$, and the
dephasing time is $T_{2}=0.0218\,ps$. The percolating cluster has been
generated inside the cube of $l_{0}=1\,\mu m$ edge having $L^{3}$ nodes, with
$L=100$ (see Fig.\ref{Pic_Fig1} (a)) that was sufficient to
simulate the percolating structure \cite{Burlak:2009a}. Each node is supposed
to indicate the position of many emitters, with the total concentration of
emitters inside the percolation cluster being $N=N_{0}+N_{1}+N_{2}%
+N_{3}=3.3\times10^{24}\,m^{-3}$. The initial (at $t=0$) values of densities
$N_{0}(0)=0.001\,N$, $N_{1}(0)=0.002\,N$, $N_{2}(0)=0.002\,N$, and
$N_{3}(0)=0.995\,N$ are used. The dielectric permittivity of a host material
is $n=2.2$ that is close to the typical values for ceramics $Lu_{3}%
Al_{5}O_{12}$, $SrTiO_{3}$, $ZrO_{2}$, see review \cite{Sanghera:2012}. The
result of simulations shown in Fig. \ref{Pic_Fig1} (b) obtained
for cw pumping given by $A_{r}=10^{7}\,s^{-1}$. At this pumping the
simulations show the formation a well-defined lasing for $t>t_{s}$, and we
refer $t_{s}$ as the lasing start time. To simulate the noise in our system
the initial seed for the electromagnetic field has been created with random
phases at each node. In what follows it is used dimensionless time $t$
renormalized as $t\rightarrow tc/l_{0}$,where $c$ is the light velocity in vacuum.

Briefly, the \textit{strategy} of our simulations consists in the following: (i)
Calculating the geometry of the spanning percolating cluster, (ii) Calculating
the photon field $\mathbf{E}$ generated by emitters incorporated in spanning
cluster with the use of finite-difference time-domain (\textit{FDTD}) technique
\cite{AllenTaflove:2005a}, and (iii) Solution of nonlinear dynamic coupled
equations for field, polarization $\mathbf{P}$ and the occupation numbers
$N_{i}$ for all the emitters in 3D system.%

We use here the extended 3D percolating model that allows the empty
percolating voids to have a small random-valued 3D shift from the reference
grid. We use the percolating probability $p=0.4$ that is slightly less of
corresponding threshold $0.415$ for such extended model.
Fig.\ref{Pic_Fig1} (a) shows the typical spatial structure of
the incipient percolating cluster in the cube $l_{0}\times l_{0}\times l_{0}$
corresponding to the numerical grid with $L\times L\times L$, $L=100$ nodes.
Fig.\ref{Pic_Fig1} (b) displays the dynamics of optical lasing
generated by exited emitters in percolating medium with intensities
$I_{x,y,z}$ through the sides of 3D sample. The normalized field amplitudes
$E_{x}$ in the central intersection ($x,y=L/2$) of 3D percolating system at
different times nearly of the initial time of laser generation ($t=4,5,6$) are
shown in Fig.\ref{Pic_Fig1}, panels (c), (d), and (e).
Fig.\ref{Pic_Fig1} (c) shows the fields distribution before the
lasing starts. In this case the field amplitudes are small, one can observe
only narrow point-like fields (indicated by arrow) generated by random
emitters in percolating cluster. At larger time $t=5$ (see
Fig.\ref{Pic_Fig1}, panel (d)) the laser generation starts that
leads to increase of the field amplitudes in the emitter positions, however
the radiation field still is small. But for $t=6$ (peak of lasing, see
Fig.\ref{Pic_Fig1}, (b), (e)) we observe the appearance of
intensive radiating light beyond the emitter peaks.%

\begin{figure}[ptb]%
\includegraphics[width=9cm]%
{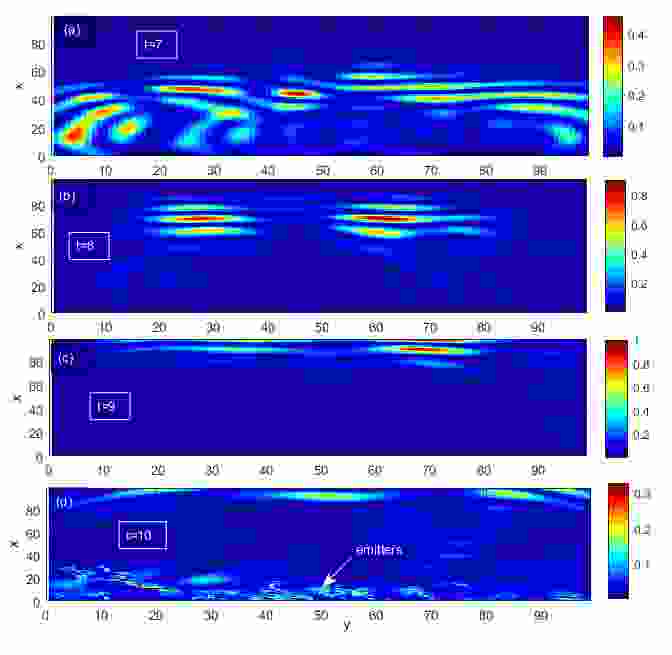}%
\caption{(Color online.) The field distribution for longer times. Panels (a)
$t=7$ and (b) $t=8$ show the dynamic states (coupled to emitters) when the
light bunches begin to migrate from the random emitter area. (That corresponds
to lasing evolution in Fig.\ref{Pic_Fig1} (b) for the interval
$t=7,8$ when in the system after emission the accumulation of energy occurs.
Panels (c) $t=9$ and (d) $t=10$ show that for long time the optical bunches
reach the output of system, and at $t=10$ (d) the localized light is radiated
to surrounding space.}%
\label{Pic_Fig2}%
\end{figure}

Fig.\ref{Pic_Fig2} shows the developed field distribution for
larger times: $t=7,8,9,10$, when the advanced lasing occurs. From this figure
one can determine $t=t_{D}$ as a maximum dwell time such that for $t\leq
t_{D}$ the dynamic bunches are still formed in all the nonlinear and
disordered part of system from $x=0$ up to $x=L/2$ and for $t>t_{D}$ the fiel
bunches are concentrated at $x>L/2$. Figures \ref{Pic_Fig2}
(a) and (b) show the nonlinear dynamics at $t=7$ and $t=8$ when well-formed
field bunches with high amplitude start to leave the random emitter area .
This corresponds to field evolution in Fig.\ref{Pic_Fig1} (b)
for $t=7,8$ when the system after emission is in the state of the energy
accumulation. Fig.\ref{Pic_Fig2} (c), (d) show that for longer
time the field bunches move to output of system; at $t=9$ they are closely to
the output and first front starts to leave the system. Moreover, we observe
from Fig.\ref{Pic_Fig2} (d) that at $t=10$ the bunches have
been radiated away the system such that the field tails have quite small
amplitude, therefore the points-like emitter fields (close to the input) are
seen in this amplitude scale. We observe that for the used parameters the
dwell time is $t_{D}\approx7$.%

\begin{figure}[ptb]%
\includegraphics[width=9cm]%
{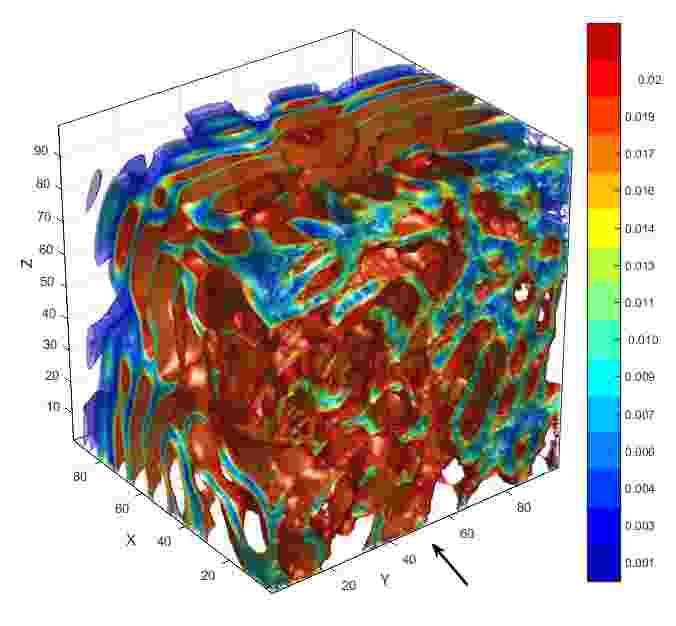}%
\caption{(Color online.) The isosurface of 3D emitted field shown in Fig.
\ref{Pic_Fig2} (a) for time $t=7$. We observe that the field
has indented shape of light structures with bunches and worm-holes due to
the fractality of radiated system.}%
\label{Pic_Fig3}%
\end{figure}

Fig.\ref{Pic_Fig2} exhibits the fields in central intersection
of 3D system. However, since the clusters have nonuniform fractal structure it
is more informative to explore the shape of fields in total 3D volume. Fig.
\ref{Pic_Fig3} displays the isosurface of field in 3D system shown in
Fig. \ref{Pic_Fig2} (a) for time $t=t_{D}=7$ when the optical
fields have already well-established structure. We observe from Fig.
\ref{Pic_Fig3} that in general field consists of bunches having
different amplitudes and shapes. The bunches are separated by 3D worm-holes
(with small field) having irregular shapes that emerge due to the non-integer
fractal dimension ($D_{H}=2.54$) of percolating clusters.

\textit{Localization.} In the above the properties of generated field bunches
were analyzed in general. In what follows we study the dynamics and structure
of localized fields confined in the light bunches. A field (quasi-spherical)
bunch will be mentioned as a bounded 3D domain where field Ei in every layer i
is a decreasing function $Ei>Ei+1$ $(i\in0,1,\ldots p-1$) from center to
periphery that leads to $E_{0}>E_{1}>\ldots>E_{p}>0$. Such domain (bunch) have
two parameters: radius $p$ and the field value in the center $E_{0}=E_{max}$.
[More detailed approaches in such 3D localization problem require too large
amount of computer resources.]%

\begin{figure}[ptb]%
\includegraphics[width=9cm]
{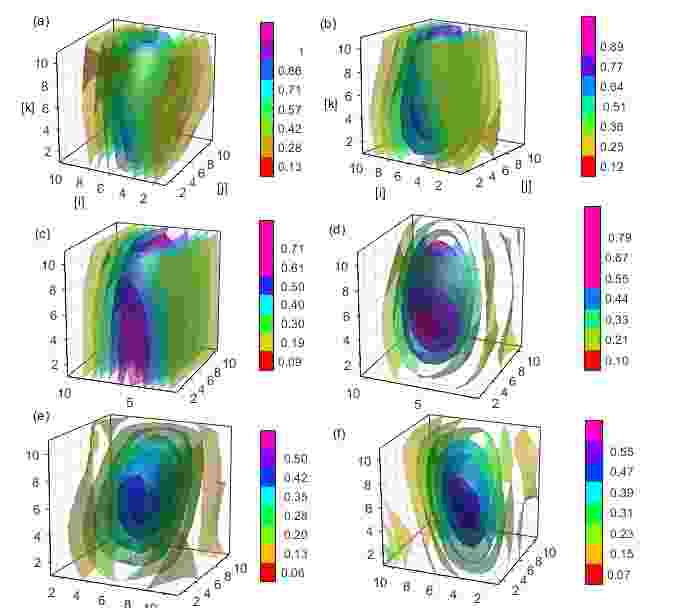}%
\caption{(Color online.) The positions and shapes of some localized field
bunches with highest (normalized) amplitudes in center $E_{max}$ and largest
radius $p$ for $t=t_{D}=7$ (see Fig. \ref{Pic_Fig2} (a))
situated in different points of 3D system: (a) $E_{max}=0.98$ at
$x=14,y=18,z=73,p=5$; (b) $E_{max}=0.96$ at $x=13,y=26,z=18,p=5$; (c)
$E_{max}=0.73$ at $x=15,y=41,z=64,p=4$; (d) $E_{max}=0.74$ at
$x=15,y=41,z=64,p=4$; (e) $E_{max}=0.5$ at $x=73,y=20,z=32,p=5$ , and (f)
$E_{max}=0.55$ at $x=18,y=35,z=21,p=5$. In general case the bunches are
situated in random positions due to disordered radiated emitters and the
fractality of percolating clusters.}%
\label{Pic_Fig4}%
\end{figure}

From Fig. \ref{Pic_Fig4} one observes for time $t=t_{D}=7$ the
structure of some localized field  with large amplitudes $E_{max}$ and
characteristic radius $p$. The centers of such objects are established in
various points of 3D medium. Fig. \ref{Pic_Fig4} displays the
isosurface of normalized fields for such items. In general the bunches are
situated in variety points of 3D system, see Fig. \ref{Pic_Fig3} due to
disordering positions of radiated emitters and the fractal shape of the
percolating clusters. As we observer from Fig. \ref{Pic_Fig4} all the
bunches have well established localized shape. %

\begin{figure}[ptb]%
\includegraphics[width=8cm]%
{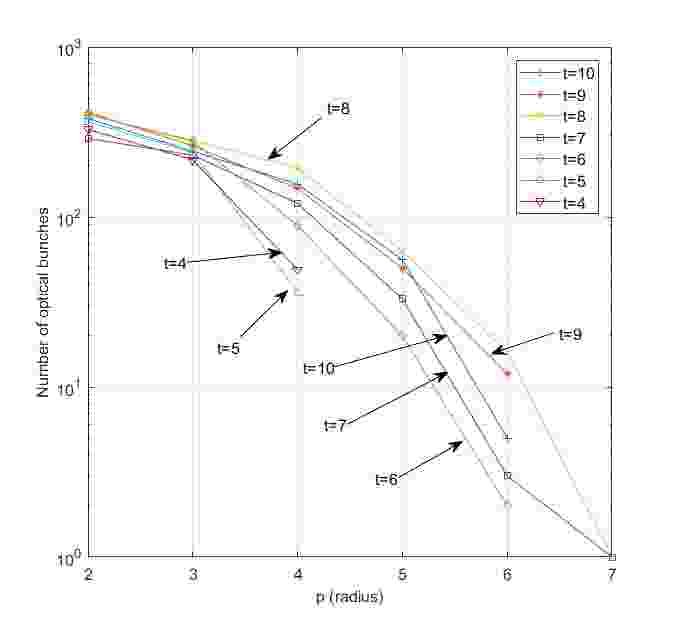}%
\caption{(Color online.) Dependence the number of optical bunches $N$ on the
radius $p$ for various times $t=4,5,6,7,8,9,10$. We observe that $N$ with
large radius $p$ drastically increments after $t>5$ when the lasing starts.
At $ t=t_{D}=7 $ and $ t=8 $ it is generated the bunch with maximal characteristic radius $ p=7$.}%
\label{Pic_Fig5}%
\end{figure}

It is interesting to study the dependence of number of bunches at different
times $t$ as function of radius $p$. Fig.\ref{Pic_Fig5}
shows such dependence for times $t=4,5,6,7,8,9,10$. We observe from
Fig.\ref{Pic_Fig5} that below the lasing threshold
($t\leq5$, see Fig. \ref{Pic_Fig1} (b)) there are large number
of bunches with small radius $p$. Most of them correspond to narrow point-like
fields generated in vicinity of random light emitters (see Fig.
\ref{Pic_Fig1} (c)). However, as one can see from
Fig.\ref{Pic_Fig5}, after the lasing starts (since $t>5$)
the number of optical bunches with large radius $p$ drastically increments.
Besides, Fig. \ref{Pic_Fig2} displays that such bunches (wth
large $p$) are situated mainly beyond the percolating area. Moreover, as Fig.
\ref{Pic_Fig2} shows such fields have well-localized shape
(Fig. \ref{Pic_Fig4}), they are not pinned to the static emitters and
migrate to the output. That allows indicating such dynamic bunches as 3D zones
of optical Anderson localization arisen by the active disordered fractal
percolating clusters.

Finally it is especially interesting to study the dependence amount of bunches at
different times $t$ as function of radius $p$.
Fig.\ref{Pic_Fig5} shows such dependence for times
$t=4,5,6,7,8,9,10$. We observe from Fig.\ref{Pic_Fig5} that
below the lasing threshold ($t\leq5$, see Fig. \ref{Pic_Fig1}
(b)) there are a lot of field peaks with small radius $p$. Most of them
correspond to narrow point-like fields generated by random light emitters (see
Fig. \ref{Pic_Fig1} (c)). However, as one can see from
Fig.\ref{Pic_Fig5}, after the lasing starts (since $t>5$)
the number of optical bunches with large radius $p$ drastically increments.
Besides, Fig. \ref{Pic_Fig2} displays that such bunches (with
large $p$) are situated mainly beyond the percolating area.  Moreover, as Fig.
\ref{Pic_Fig2} shows such fields have well-localized shape
(Fig. \ref{Pic_Fig4}), they longer are not pinned to the static
emitters and they migrate to the output of system. That allows indicating such
dynamic bunches as 3D zones of optical Anderson localization arisen by the
active fractal percolating clusters.

\textit{Conclusion.} We have shown that dynamic 3D optical Anderson
localization can arise in medium with percolating disorder when the
percolating clusters are filled by the light nanoemitters in the excited
state. In such materials the field emitters in clusters create a fractal
radiating system, where the field is not only emitted, but also scattered by
the boundaries of clusters. Our 3D simulations have found that in such
nonlinear compound the amount of localized field bunches drastically increases
after the optical lasing starts. At longer times all the bunches leave
nonlinear percolating structure and are radiated from the medium.


\begin{thebibliography}{99} 
	                                                       
\bibitem {Riboli:2014}F. Riboli, N. Caselli, S. Vignolini, et.al., Nature
Materials, Vol. 13 (7), pp. 720-725 (2014).

\bibitem {Vinck:2011}K. Vynck, M. Burresi, F. Riboli, D. S. Wiersma, Nature
Materials, vol. 11, pp. 1017--1022 (2012).

\bibitem {Sheng:2010}P. Sheng, Introduction to Wave Scattering, Localization
and Mesoscopic phenomena (Springer, 2010), 2nd ed.,.

\bibitem {Wang:2011}J. Wang and A. Z. Genack, Nature 471, 345 (2011).

\bibitem {jendrzejewski:2012}F. Jendrzejewski, A. Bernard, K. Muller, P.
Cheinet, et. al., Nature Physics 8, 398--403 (2012).

\bibitem {Segev:2013a}M. Segev, Y. Silberberg, D. N. Christodoulides, Nature
Photonics 7, 197--204 (2013)

\bibitem {Wiersma:2013a}D. S. Wiersma, Nature Photonics. 7, 3, 188-196 (2013).

\bibitem {anderson:1958}P. W. Anderson, Phys. Rev.,109, 1492 (1958).

\bibitem {Skipetrov:2016a}Skipetrov, S.E., Phys. Rev. B. 94, 6, 064202 (2016).

\bibitem {Sanli:2009a}Sanli Faez, Anatoliy Strybulevych, John H. Page, Ad
Lagendijk, and Bart A. van Tiggelen, Phys. Rev. Lett. 103, 155703 (2009).

\bibitem {Burlak:2009a}G. Burlak, M. Vlasova, P. A. Marquez Aguilar, et.al.,
Opt. Commun. \textbf{282}, 2850 (2009).

\bibitem {burlak:2015}G. Burlak and Y. G. Rubo, Phys. Rev. A, 92, 013812, 2015.

\bibitem {Sanghera:2012}J. Sanghera, W. Kim, G. Villalobos, et.al., Materials
\textbf{5}, 258 (2012).

\bibitem {Siegman:1986}A. E. Siegman, \textit{Lasers} (Mill Valley,
California, 1986).

\bibitem {Soukoulis:2000}Xunya Jiang and C. M. Soukoulis, Phys. Rev. Lett.
\textbf{85}, 70 (2000).

\bibitem {conti:2008a}C. Conti and A. Fratalocchi, Nat. Phys. \textbf{4}, 794 (2008).

\bibitem {AllenTaflove:2005a}A. Taflove and S. C. Hagness, \emph{Computational
Electrodynamics: The Finite-Difference Time-Domain Method} (Artech House
Publishers, 3rd edition, 2005).

\bibitem {expt}H. Cao, Y. G. Zhao, S. T. Ho, E. W. Seelig, Q. H. Wang, and R.
P. H. Chang, Phys. Rev. Lett. \textbf{82}, 2278 (1999).

\bibitem {Billy:2008a}J. Billy, V. Josse, Z. Zuo, et.al. Nature 453, 891-894 (2008)
\end{thebibliography}
\end{document}